\documentclass[aps,prl,reprint,groupedaddress]{revtex4-1}

\usepackage[dvips]{graphicx}
\usepackage{amssymb,amsfonts,amsmath}
\usepackage{natbib}
\usepackage{color}

\def\half{{\textstyle \frac{1}{2}}}
\def\quarter{{\textstyle \frac{1}{4}}}

\begin{document}


\title{Growth and Shape of a Chain Fountain}


\author{John S Biggins}
\affiliation{Cavendish Laboratory, University of Cambridge, Cambridge, United Kingdom}


\date{\today}
\begin{abstract}If a long chain is held in a pot elevated a distance $h_1$ above the floor, and the end of the chain is then dragged over the rim of the pot and released, the chain flows under gravity down into a pile on the floor. Not only does the chain flow out of the pot, it also leaps above the pot in a ``chain-fountain''. I predict and observe that the steady state shape of the fountain is an inverted catenary, and discuss how to apply boundary conditions to this solution. In the case of a level  pot, the fountain shape is completely vertical. In this case I predict and observe both how fast the fountain grows to its steady state height, and how it grows $\propto t^2$ if there is no floor. The fountain is driven by an unexpected push force from the pot that acts on the link of chain about to come into motion. I confirm this by designing two new chains, one consisting of hollow cylinders threaded on a string and one consisting of heavy beads separated by long flexible threads. The former is predicted to produce a  pot-push and hence a fountain, while the latter will not. I confirm these predictions experimentally. Finally I directly observe the anomalous push in a horizontal chain-pick up experiment.\end{abstract}



\maketitle


The mechanics of chains is one of the oldest fields in physics. Galileo observed that hanging chains approximate parabolas, particularly when the curvature is small\cite{galilei1974two}, while the true shape was proved to be a catenary by Huygens Leibniz and John Bernoulli\cite{lockwood1971book}.  
A chain hanging in a catenary is a structure supporting its weight with pure tension. In 1675 Hooke discovered that  a thin arch supporting its own weight with pure compression  must follow the inverted shape of a hanging chain\cite{hookedescription}, that is, an inverted catenary. Ever since architects from Wren to Gaudi have incorporated inverted catenary arches into their buildings and even used hanging strings to build inverted architectural prototypes. We might expect such a venerable and technologically important field to  have few remaining surprises, but chain mechanics has recently produced several. A chain falling onto a table accelerates faster than $g$, leading inexorably to the conclusion that the table must pull down on the falling chain\cite{grewal2011chain, hamm2010weight}. If a pile of chain rests on a surface, and the end is then pulled in the plane of the surface to deploy the chain, an unexpected noisy chain arch has been observed to form perpendicular to the surface of the chain immediately beyond of the pile\cite{Santagelochainarch}, that is, in the portion of chain that has just come into motion. There is also recent work on the rich dynamics of whips and free ends\cite{shapeofawhip,HannaSantangelofreeend,tomaszewski2006motion}.

\begin{figure}
\centerline{\includegraphics[width=0.3\textwidth]{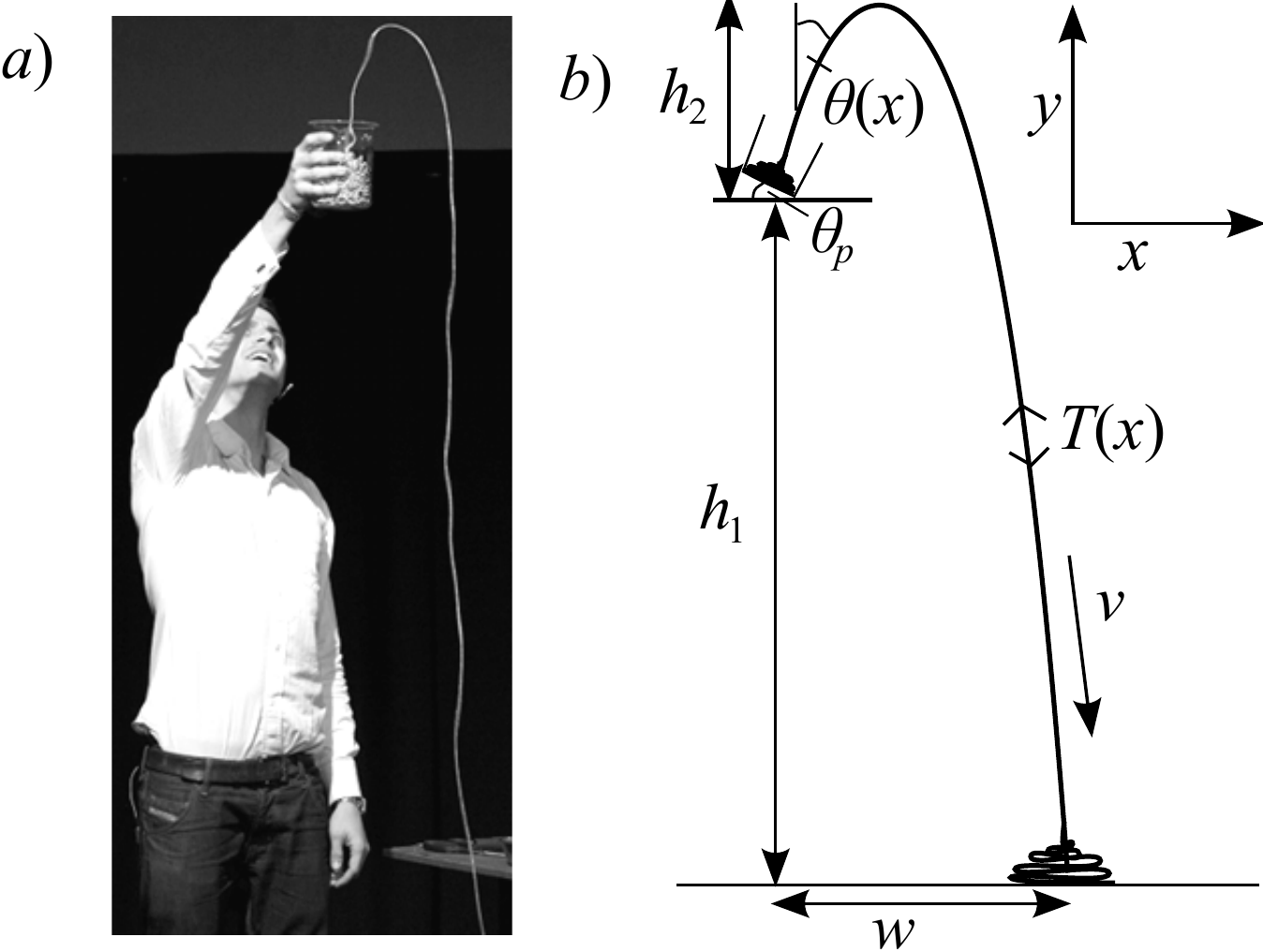}}
\caption{a) Steve Mould demonstrating a chain fountain. Photo courtesy of J.\ Sanderson. b) Diagram of a chain fountain. A chain with mass per unit length $\lambda$ flows at speed $v$ along a curved trajectory from a pot tilted to an angle $\theta_p$ and elevated to an height $h_1$, to the floor. The fountain has height  $h_2$ and width $w$. At each point $x$ the chain has a height $y(x)$ a tension $T(x)$ and makes an angle $\theta(x)$ with the vertical.}\label{photoanddiagram}
\end{figure}

The most recent surprise comes via Mould's videos of a chain fountain\cite{mouldwebsite}, shown in fig.\  \ref{photoanddiagram}a, in which a chain not only flows from an elevated pot to the floor under gravity but leaps above the pot. These videos have surprised and delighted almost 3.5 million viewers. In this letter I demonstrate that chain in such a fountain traces Hook's inverted catenary, but as a structure of pure tension stabilized by the motion of the chain. 

In a chain fountain, the leaping of the chain above the pot requires that when a link of the chain is brought into motion, it must not only be pulled into motion by the moving chain but also pushed into motion by the pot\cite{BigginsWarnerChain}.  This anomalous push is expected to arise whenever a pile of chain is deployed and, as such, has a wide field of potential applications. However, the analysis in \cite{BigginsWarnerChain} infers the existence of the anomalous push from a simplified model of a  zero-width steady-state fountain, leading to questions about whether the anomalous push is an artifact of these assumptions.  In this letter I consider fountains of finite width and the dynamics of fountain growth. The extended theory does not mitigate the need for an anomalous force, and explains the observed fountain behavior well. I also confirm the anomalous force hypothesis experimentally, both by direct observation in a horizontal pickup geometry, and by comparing the fountains made by radically different sorts of chain.


A non-vertical chain fountain is sketched in fig.\ \ref{photoanddiagram}b. We expect that, after the fountain reaches the floor, it will tend to a steady shape.To find this equilibrium curve, consider an element of chain with  horizontal extent $\mathrm{d}x$, which  has  length $\mathrm{d}s=\mathrm{d}x/\sin{(\theta)}$ and  mass $\lambda \mathrm{d}s$. Tangentially there is no acceleration so the tension gradient  balances gravity, 
\begin{equation}
T'(x)=\frac{\lambda g}{\sin{\theta}} \cos{\left(\theta\right)} .
\end{equation}
Since $\cot{\left(\theta\right)}=y'(x)$ this can be  integrated to give
\begin{equation}
T(x)=\lambda g y +\lambda(v^2-c g),
\end{equation}
where we have written the constant of integration as $\lambda(v^2-c g)$ and $c$ is a constant. Perpendicularly, there is the inward force $T(x)/r(x)$  (where $r(x)$ is the radius of curvature), a Laplace-pressure like term that arrises whenever one has tension in a curved surface. This force and gravity supply the  centripetal acceleration:
\begin{equation}
\frac{T(x)}{ r(x)}-\lambda g \sin{\left(\theta\right)} =\lambda \frac{v^2}{r(x)}.
\end{equation}
Recalling that in Cartesians $1/r=y''(x)/(1+y'(x)^2)^{3/2}$ and $\sin{\left(\theta\right)}=1/\sqrt{1+y'(x)^2}$, this simplifies to
\begin{equation}
(T(x)-\lambda v^2)y''(x)=g \lambda (1+y'(x)^2).\label{cateqn1}\end{equation}
Substituting in our result for $T(x)$ and solving for $y(x)$ reveals that a chain moving along its own length under gravity in an unchanging shape must trace a catenary.\cite{Tripos_1854,airy1858mechanical,perkins1989theoretical}.  Curiously, this result was first published as a question in the 1854 Cambridge University maths examination\cite{Tripos_1854}.  In the case of the fountain, this catenary must be an inverted one, viz.
\begin{equation}
y(x)=-a \cosh{\left(\left(x-b\right)/a\right)}+c
\end{equation}
where $a$, $b$ are new constants of integration.

\begin{figure}\centering
\includegraphics[width=0.45 \textwidth]{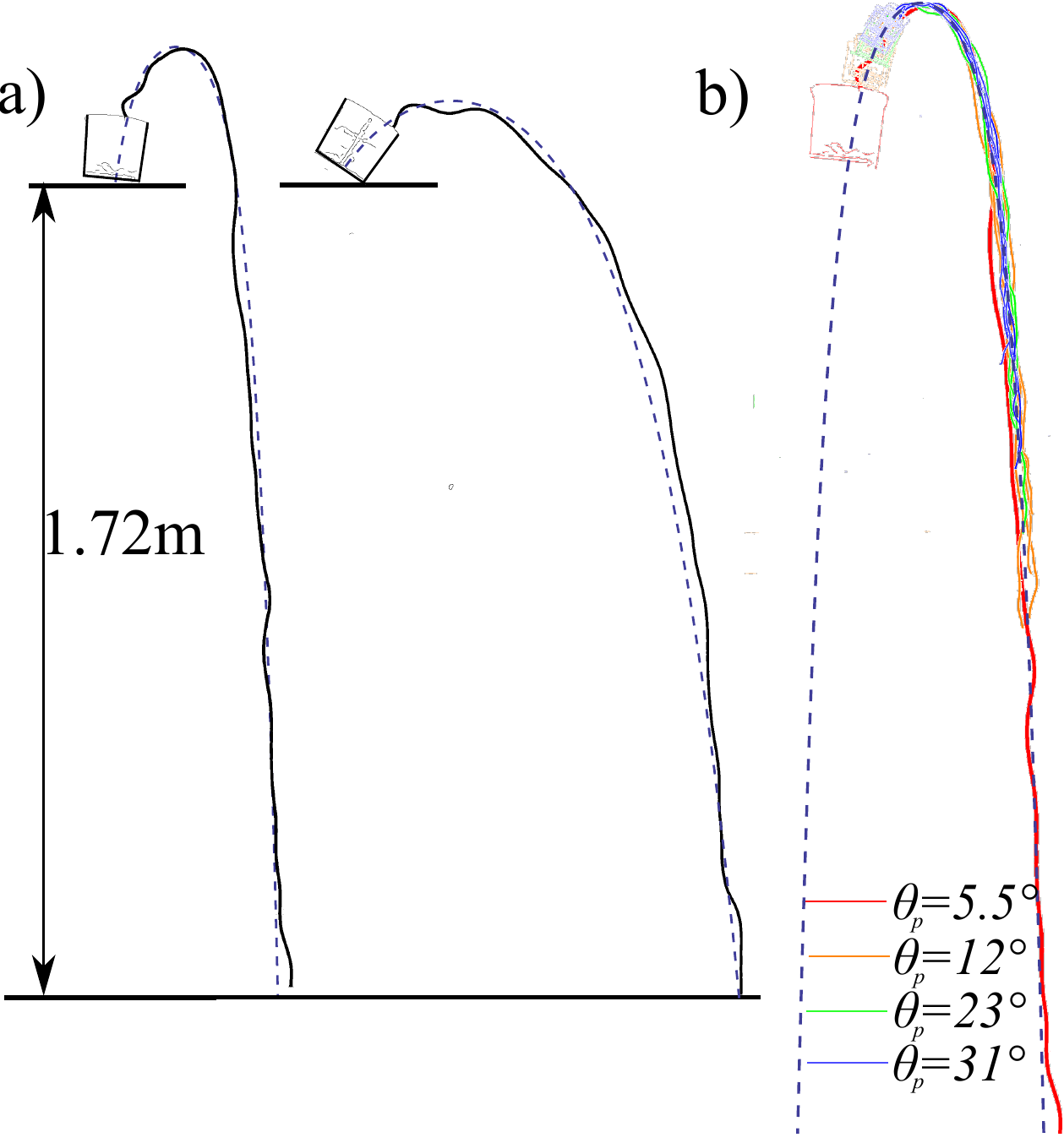}
\caption{a) Two experimental chain fountains (black lines), both  obtained with a drop $h_1$ of 1.72m but with different pot tilt angles (left: $\theta_p=5.5^{\circ}$ right: $\theta_p=31^{\circ}$) leading the fountains to have different widths. Both fountains are fitted well by a scaled catenary (blue dashed line) b) Instances of many chain fountains, coloured according to pot tilt angle, scaled and superimposed so that they all collapse onto a single catenary shown with a blue dashed line.} \label{chaincats}
\end{figure}

The simplest  inverted catenary is  $y(x)=-\cosh{(x)}$.  The above solution is simply this curve translated and with a ``zoom'' by a factor of $a$ equal to the radius of curvature of the catenary at its apex.  All steady-state chain fountains should therefore produce shapes that, after  zooming and translating,  collapse onto $-\cosh(x)$. To test this, a 50m long  brass ball-chain was put in a 1L  beaker, elevated to 1.72m above the ground and tilted by an angle $\theta_p$. The end of the chain was then pulled over the rim  and released initiating a chain fountain. The experiment was repeated with different tilt angles, resulting in different fountain shapes, which were photographed towards the end of each run to ensure the fountain was in its steady state. Runs with significant tangles were disregarded. Two examples are shown in fig.\  \ref{chaincats}a, one thin and one wide.  The fountains undulate locally but macroscopically trace a catenary.  In fig.\  \ref{chaincats}b many chain fountains with different widths are rescaled onto a single inverted catenary, demonstrating that  the chain fountain is  well described by Hook's inverted catenary.

To determine the  parameters, $a$, $b$ and $c$, $v$ and $w$ (the width of the fountain, see fig.\ \ref{photoanddiagram}b) we require five boundary conditions. The first two, $y(0)=0$, and $y(w)=-h_1$, fix the coordinate origin and the fountain drop. To find the remaining three  we must examine the pickup and putdown processes.

In a time $\mathrm{d}t$ a length of chain $v\mathrm{d}t$ is picked up, acquiring a momentum $\lambda v^2 \mathrm{d}t$. If the links are  accelerated solely by the tension, then the third boundary condition is $T(0)=\lambda v^2$. However, inspecting eqn.\ (\ref{cateqn1}), we see that the right-hand side is always finite so setting $T(0)=\lambda v^2$ require $y''(0)\to \infty$, corresponding to the chain reversing direction immediately above the pot. For a fountain we must have $T(0)<\lambda v^2$. However, the total force must still be $\lambda v^2$, so we must introduce an anomalous reaction force from the pot pushing the links into motion $f_p=\alpha \lambda v^2$, reducing the tension to $T(0)=(1-\alpha)\lambda v^2$.
\begin{figure*}
\includegraphics[width= \textwidth]{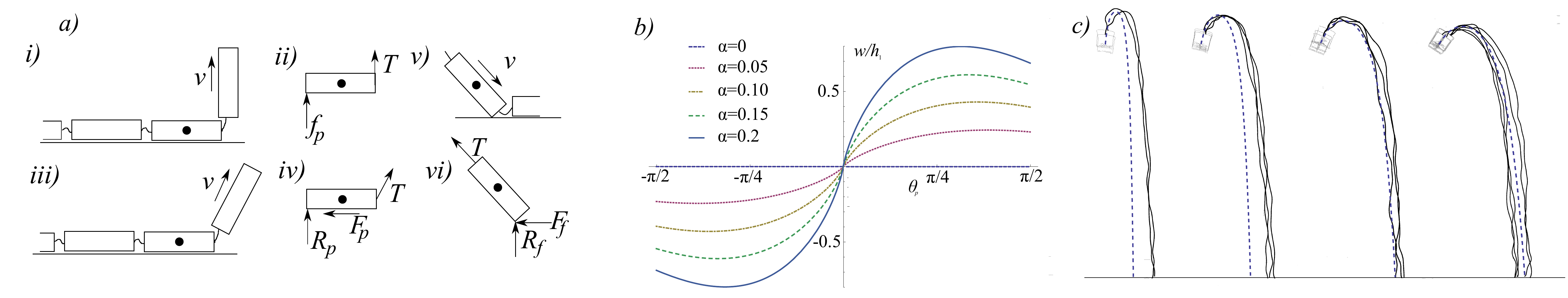}
\caption{\label{fig:wide}a.i) Sketch of a chain of rigid rods being picked up vertically from a surface. a.ii) Force diagram of the rod about to be picked up, including a tension $T$ from the moving part of the chain and a reaction $f_p$ from the surface. a.iii-iv) Same as a.i-ii but with a non-perpendicular pickup. The surface can now respond with a reaction force $R_p$ and a frictional force $F_p$. a.v) A chain being deposited on a surface at an angle. a.vi) Force diagram of a link being brought to a halt but tension and reaction and frictional forces from the floor. (b) Prediction for the width of the fountain $w$, in units of the fountain drop $h_1$, plotted against pot tilt angle $\theta_p$. The lines correspond to different anomalous pushes from the pot during parameterized by $\alpha$. If $\alpha=0$  then $w=0$. The plot is drawn with  $\beta=0.2$. Reducing $\beta$ only has only a modest effect on the curves: e.g.\  $\beta=0$   reduces the curves to about two thirds of the shown height (c)Experimental snapshots of chain fountains with pot tilt angles of (left to right) $5.5^{\circ}$, $12^{\circ}$, $23^{\circ}$ and $31^{\circ}$ and a drop of 1.72m overlaid with the predicted catenary shape using $\alpha=0.12$ and $\beta=0.11$ in a blue dashed line.}\label{widefig}
\end{figure*}

We understand this anomalous pot push by modeling the chain as freely jointed rigid rods being picked up vertically from a horizontal surface, fig.\  \ref{widefig}ai. The next rod is pulled upward into motion by  a tension force $T$  at its end, sketched in fig.\  \ref{widefig}aii, causing the center of mass of the rod to lift and the rod to rotate. In isolation this would cause the far end of the rod to move down. In reality the surface prevents this  by pushing up with a reaction force $f_p$. We can estimate the size of this force, and hence $\alpha$, by considering the forces on the rod at the first moment of pickup when the rod is still horizontal. The initial linear and angular accelerations of the rod are  $m \dot{v}=T+f_p$ and  $I \dot{\omega}=(f_p-T)l/2$, where $v$ is the velocity of the rod's center of mass, and $I$ and $\omega$ are the rod's moment of inertia and angular velocity,  both around the center of mass. The initial acceleration of the left hand tip of the rod must be zero, requiring $\dot{v}+\dot{\omega}l/2=0$.  Writing $f_p=\alpha \lambda v^2$ and $T=(1-\alpha)\lambda v^2$ we can rearrange this to estimate $\alpha$ as
\begin{equation}
\alpha=\half\left(1-I/\left(\quarter m l^2\right)\right).\label{alphaeq}
\end{equation}
This calculation is not intended as a precise calculation of the value of $\alpha$, but rather to illustrate  how a surface can push on a departing chain, and hence motivate taking a non-zero value for $\alpha$ at all.  This simple calculation  demonstrates that the pot can push, justifies the functional form $f_p=\alpha \lambda v^2$  and reveals that $\alpha$ depends on the details of the chain, here via the ratio $I/(m l^2)$. A better estimate would require us to include angle fluctuations in both the pile of rods and the departing chain, and  to consider the complete pickup-process rather than just the first moment. We can use the above form to crudely estimate $\alpha$ for the ball-chain used in this paper. It takes 6 balls for the chain to turn $180\,^{\circ}$ so, in a freely jointed rod-model, the rod-like unit must be  three balls connected by two rods. Treating the balls as point masses and the rods as light, this gives $I=2(m/3)(l/2)^2=m l^2/6$, and hence $\alpha=1/6=0.166..$ which, as expected, is close to but somewhat larger than the best-fit experimental value of $\alpha=0.12$.

If the chain departs at an angle, sketched in fig.\  \ref{widefig}aiii, the  velocity and tension  point along the  chain's tangent. Momentum conservation requires any force from the surface to point in the same tangential direction and hence have an in-surface component which can only arise as a frictional force. Since  friction always acts to hinder or prevent motion, it will always oppose the in-plane component of the tension, shown in fig.  \ref{widefig}aiv,  the total surface force will only point in the tangential direction if pickup is perpendicular to the surface, giving a fourth boundary condition $\theta(0)=\theta_p$.
 
At the floor a  force $\lambda v^2$ is required to halt the  chain. If all this force comes from the floor then the tension at the end is $T(w)=0$. However, careful observations of free-falling chains dropping onto hard surfaces reveal that they accelerate faster chains that fall freely \cite{grewal2011chain, hamm2010weight}. This demands $T(w)>0$ so that the  force on the falling chain is greater than its weight, so we set $T(w)=\beta \lambda v^2$. If the chain hits the floor at an angle, the friction and reaction from the floor are both opposite to the chain's velocity and can add to point along the chain's tangent, as sketched in fig.\  \ref{widefig}v-vi. There is thus is no constraint on $\theta(w)$. 

The five boundary conditions
\begin{align}
 y(0)&=0\hspace{2em} \theta(0)=\theta_p\hspace{2em}T(0)=(1-\alpha)\lambda v^2\notag \\
y(w)&=-h_1\hspace{ 2 em} T(w)=\beta \lambda v^2,
\end{align}
determine the five unknowns in the catenary solution as
\begin{align}
a&=\frac{\alpha  h_1 \sin (\theta_p)}{1-\alpha -\beta }\hspace{0.5em}c=\frac{\alpha  h_1}{1-\alpha -\beta }\hspace{0.5em}v=\sqrt{\frac{g h_1}{1-\alpha -\beta }}\\
b&=a \sinh ^{-1}(\cot (\theta_p))\hspace{1.5em}w=a\cosh ^{-1}\left((c+h_1)/a\right)+b.\notag
\end{align}
 
We see the chain's trajectory is independent of $g$ though the chain's speed is not. We plot the fountain width, $w$, as a function of $\theta_p$ in fig.\  \ref{widefig}b. We see that for small tilt angles the width of the fountain is zero, but that the gradient $\mathrm{d}w/\mathrm{d}\theta_p$ is divergent, albeit weakly as $\sinh^{-1}(\cot(\theta_p))\sim\log(\theta_p)$, and that for large tilt angles the fountain width starts to decrease.

A unit length of chain releases gravitational energy $g h_1 \lambda$ and receives kinetic energy $\half g h_1 \lambda/(1-\alpha-\beta)$. In the traditional regime ($\alpha=\beta=0$) half the gravitational energy is thus lost in the chain pickup process, as classically expected when picking up a chain at constant velocity\cite{BigginsWarnerChain}. The additional anomalous forces reduce this energy loss. The requirement that the kinetic energy not be larger than the released gravitational potential energy imposes the bound $\alpha+\beta<\half$.

The fountain height above the pot is simply $h_2=y(b)$:
\begin{equation}
h_2=\frac{\alpha  h_1 (1-\sin (\theta_p))}{1-\alpha -\beta }.
\end{equation}
Both the height and the width of the fountain vanish if $\alpha=0$ (but not if $\beta=0$) establishing that the push from the pot during the chain pickup is the driver of a chain fountain. If the pot is level ($\theta_p=0$) the height is maximal and $w=0$ so the fountain is vertical.

We compare the predicted and experimental shapes of chain fountains in fig.\  \ref{widefig}c. Fixing $h_1$  and  $\theta_p$ to their experimental values, a good match is achieved for large-angle fountains, and the height of all fountains, by taking  $\alpha=0.12$ and $\beta=0.11$, consistent with \cite{grewal2011chain} and \cite{BigginsWarnerChain}. However, although the small angle fountains are  catenaries, the predictions are too thin. This is true for all physical values of  $\alpha$ and $\beta$ so we conclude that $\theta(0)$ is actually slightly larger than $\theta_p$, and thus that the argument for perpendicular pickup must be lacking. This may be because it neglects the pot's finite width, which for small $\theta_p$ is comparable to the pot-level width of the catenary, leading to substantial variation in the pickup angle as the pickup point move around the pot. It may also be because the rods do not lie flat on a flat pot surface, but at  at somewhat random angles on a rough bed of other rods, so the rods can experience non-frictive reaction forces that are not  perpendicular to the pot base, or because the rods come into motion prior to being picked up.  Divergence of $\mathrm{d}w/\mathrm{d}\theta_p$ for $\theta_p\to0$  makes the system sensitive to any deviation of $\theta(0)$ from $\theta_p$ at small angles, where the discrepancy between actual and predicted width is seen.

\begin{figure}\centering
\includegraphics[width=0.48 \textwidth]{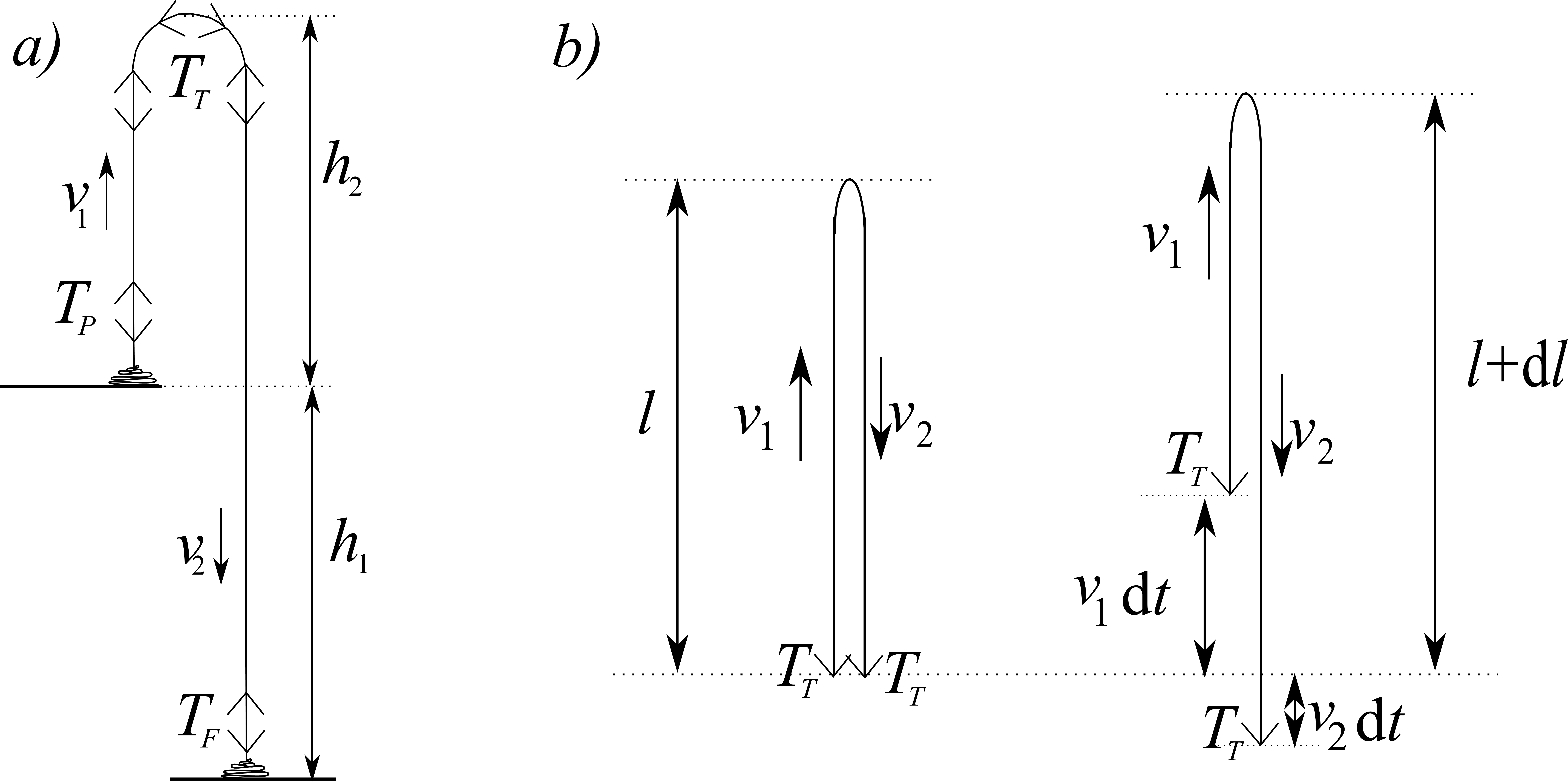}
\caption{a) Diagram of a growing vertical chain fountain. While the fountain grows the velocities of the two legs are not equal. b) Diagram of a small length ($2l$) of chain flowing around the apex of the fountain. The same material is shown on the left, and a short time $\mathrm{d}t$ later on the right. }\label{vertfountain}
\end{figure}

We now consider the growth of the fountain. We focus on the case of a level pot where the fountain is vertical. We sketch such a fountain in fig.\  \ref{vertfountain}a, labeling the speeds of the rising and falling legs   $v_1$ and $v_2$. As before, the tension above the pot is $T_P=(1-\alpha )\lambda v_1^2$ and that above the floor is $T_F=\beta\lambda v_2^2$. Fig.  \ref{vertfountain}b shows two diagrams, separated by a time $\mathrm{d}t$, of a section of chain, length $2l$, traversing the apex. Inextensibility requires $2 \mathrm{d}l+ v_2\mathrm{d}t=v_1\mathrm{d}t$. However, $\mathrm{d}l/\mathrm{d}t=\dot{h}_2$, so this constraint gives
\begin{equation}
\dot{h}_2=\half\left(v_1 - v_2\right).\label{h2d}
\end{equation}
Further, a length  $v_1\mathrm{d}t-\mathrm{d}l=\half(v_1+v_2)\mathrm{d}t$ is converted from moving upward at $v_1$ to downward at $v_2$, requiring a momentum change of  $\mathrm{d}p=\half\lambda \left(v_1+v_2\right)^2\mathrm{d}t$. This momentum is provided by the tension $T_T$ which acts on both sides of the apex, requiring $2 T_T=\half \lambda \left(v_1+v_2\right)^2.$ Momentum balance in the two legs then requires
\begin{align}
- g h_2+\quarter  \left(v_1+v_2\right)^2-(1-\alpha) v_1^2&=h_2 \dot{v}_1\label{eqn3}\\
- g (h_1+h_2)\hspace{-0.1 em}+\quarter  \left(v_1+v_2\right)^2-\beta  v_2^2&=-(h_1+h_2)\dot{v}_2.\label{eqn4}
\end{align}

\begin{figure}\centering
\includegraphics[width=0.45 \textwidth]{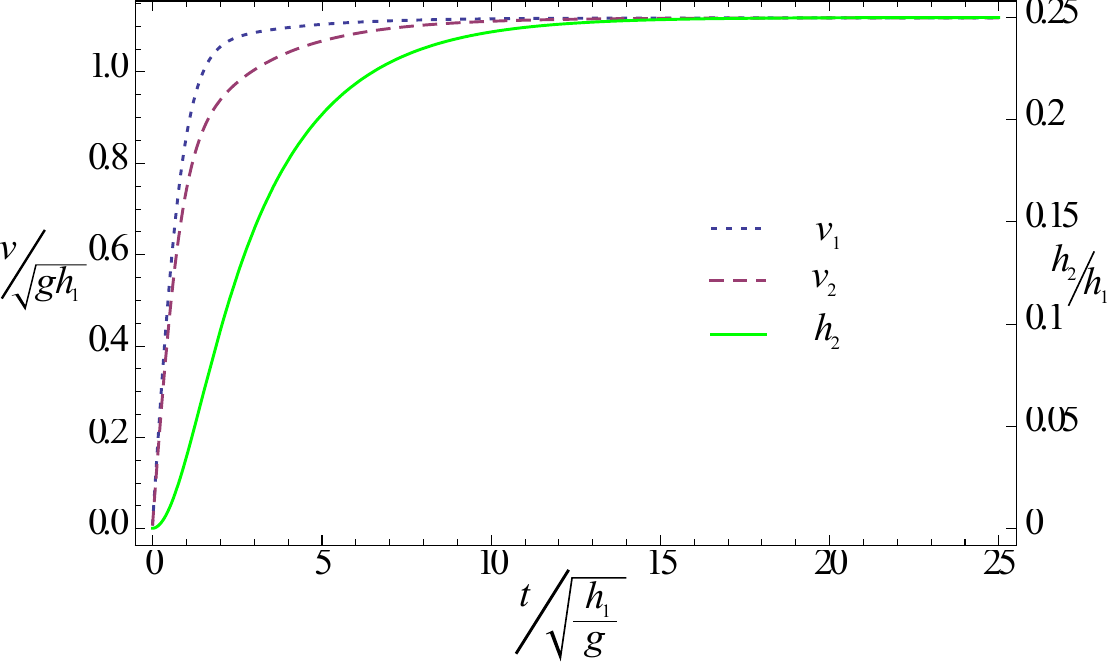}
\caption{Numerical solution to eqns.\ (\ref{eqn3}-\ref{eqn4}) calculated using $\alpha=0.2$ and $\beta=0$. The velocities of both legs of the fountain ($v_1$ and $v_2$) and the fountain height rise from zero to their (equal) steady state values in a timescale comparable to  $\sqrt{h_1/g}$.} \label{examplenumintegrate}
\end{figure}

\begin{figure}\centering
\includegraphics[width=0.48 \textwidth]{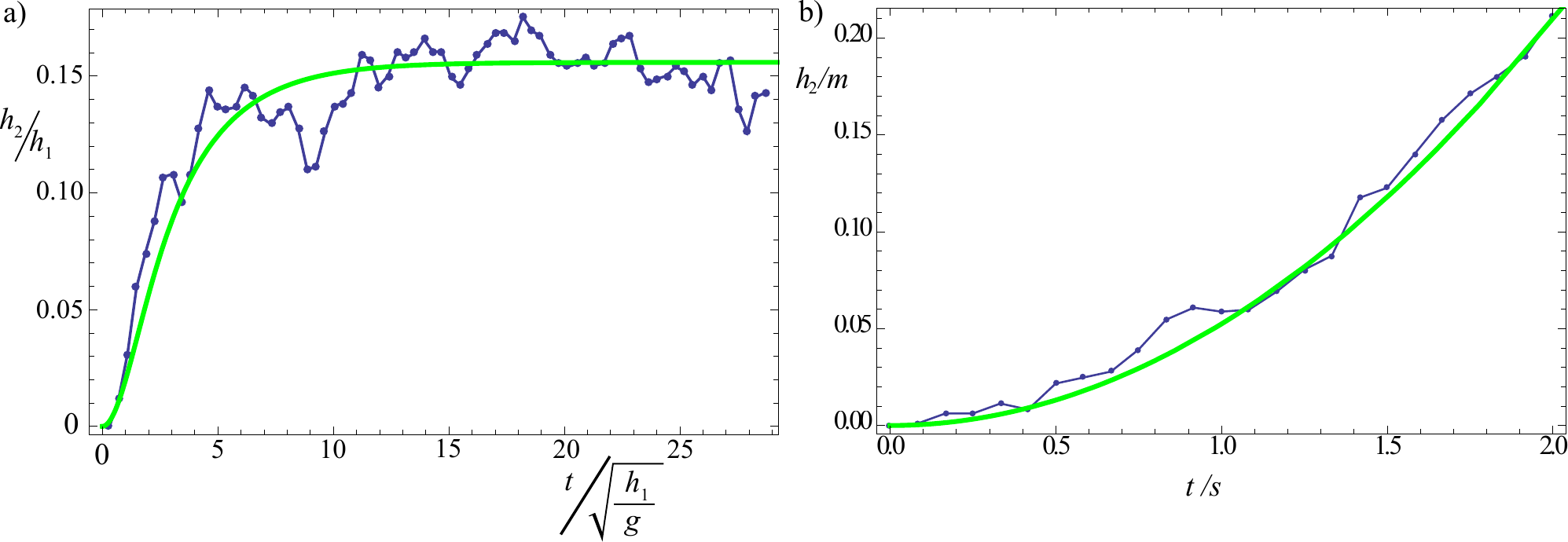}
\caption{ Comparisons of experimentally observed  (joined blue dots) and theoretically predicted (green line) fountain height ($h_2$) as a function of time, $t$. a) Experiment using a drop of 1.8m showing a steady state height and a theoretical line calculated using $\alpha=0.12$ and $\beta=0.11$. b) Experiment with no floor showing quadratic growth of the fountain height. Theoretical line calculated using $\alpha=0.12$. } \label{expgrowthgraphs}
\end{figure}

The dynamic equations (\ref{h2d}-\ref{eqn4}) are not analytically integrable, though in the steady state they reproduce the earlier analysis. Dimensional analysis reveals the fountain must grow to a height proportional to $h_1$ in a time proportional to $\sqrt{h_1/g}\sim 0.5\mathrm{s}$ . We confirm this via numerical solutions,  shown in fig.\  \ref{examplenumintegrate},  obtained using the initial conditions $h_2=v_1=v_2=0$ at $t=0$.  Experimental data for fountain growth was obtained, again using a $1.8$m drop,  by recording the height of the apex above the level of beads in the pot every 0.17s. The pot was tilted to a small angle $\theta_p\sim 1.5^{\circ}$ to ensure the chain exited on one side. The fountain was initiated with the chain  touching the ground to match the theory. The total length of the chain was 50m, and the fountain lasted 12s. We compare this data  with the theoretical prediction in fig.\  \ref{expgrowthgraphs}a, using the same chain parameters ($\alpha=0.12$, $\beta=0.11$) as in the catenary analysis. We see that the experimental line is quite noisy, but the theoretical line, with no new fitting parameters, matches well.

If there is no floor the end of the chain falls ever lower. The relevant equations are still eqns (\ref{h2d}-\ref{eqn4}) but are augmented by $\dot{h}_1=v_2$, and modified by setting $\beta=0$ since, as there is no floor, it cannot provide any  force. These new equations  admit a continuously growing analytic solution with $v_1\propto v_2\propto gt$ and $h_1\propto h_2\propto g t^2$, where the constants of proportionality only depend on $\alpha$. In particular,
\begin{align}
h_2=\frac{\left( 4 \alpha +3 \sqrt{4-3 \alpha }-6 \right) g t^2}{42-32 \alpha }.
\end{align}
We test this in fig.  \ref{expgrowthgraphs}b using an 8m drop, provided by a 3-story stairwell in the Canvendish Laboratory, and an 8m length of the same ball chain. We again achieve a good fit using $\alpha=0.12$.

\begin{figure}\centering
\includegraphics[width=0.48 \textwidth]{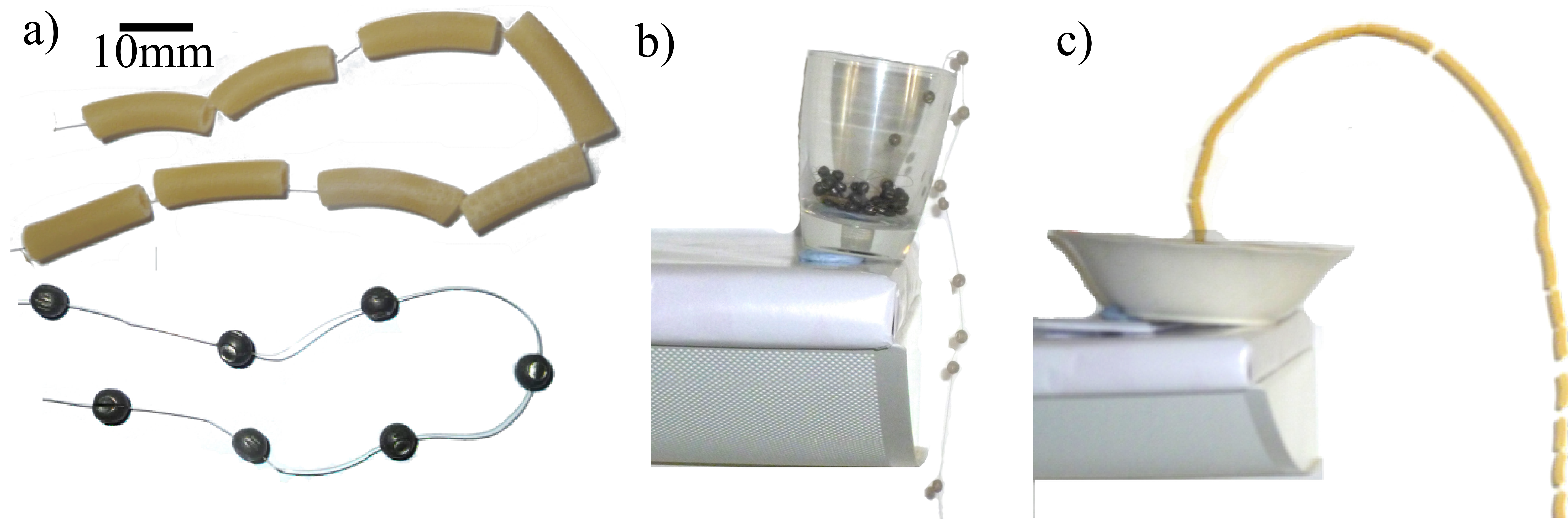}
\caption{Experimental chain fountains with two very different sorts of chains. a) Image of the two chains. The top chain consists of hollow cylinders threaded on a nylon thread. The lower chain is heavy tungsten beads separated by long stretches of fine thread. b) The bead-chain does not produce a fountain with a 1.8m drop. c) The cylinder chain does produce a fountain with a 1.8m drop.}\label{chaintests}
\end{figure}

The driver of the chain fountain is the surprising extra push from the pot that helps launch links of the chain into motion. We test the suggested origin of this force by considering two very different chains. The first consists of small (diameter 4mm) spherical tungsten beads separated by 2.5cm lengths of  thread. This does not produce a push since the beads are picked up individually and their round shape means any rotation induced does not result in them pushing down, so this chain should not produce a fountain. The second chain consists of short hollow cylinders (in reality pieces of uncooked macaroni pasta,  2cm long and 4mm  diameter) strung end to end on fine thread. It resembles the chain of freely-jointed rods  used to derive eqn.\ (\ref{alphaeq}), and hence should produce a reaction and a fountain. The chains, shown in fig.\  \ref{chaintests}a,  had similar mass densities, were 10m long and were dropped from a pot 1.8m above the ground. The fountains produced are shown in figs.\ \ref{chaintests}b-c. The result that the bead chain does not produce a fountain while the freely-joined-rod chain does is unambiguous.

To directly observe the push, 8m of ball-chain was  closely packed in serpentine rows on a table, and the end was then released over the end of the table, causing the chain to deploy horizontally  perpendicular to the rows. As first reported online\cite{Geminardtablefall}, and seen in supplementary video 1, during the experiment the rows of chain moves backwards, implying that they push forwards on the departing chain.

Observation of the chain fountain's catenary shape and its growth and saturation reveal that, although it fluctuates considerably, it does so around the steady state discussed in this paper. However, the angle the chain fountain leaves a slightly tilted pot remains unquantified and the noisy nature of the fountain  is  poorly understood. The energetic  origin of the noise is clear --- it comes the gravitational potential energy  that is ``dissipated'' during the pick-up process --- however, its amplitude and wavelength are unquantified. These questions probably relate to the finite pot width and the chaotic ordering of the chain pile, suggesting further work varying pot width and using ordered  piles. The relationship between $\alpha$  and $\beta$ also merits further work. They are very similar for the ball chain, leading one to wonder whether they are always equal. 

This work also complicates the idea of a perfectly flexible string. Eqn.\ (\ref{alphaeq}) shows that  $\alpha$ for a chain of rods depends only on $I/(ma^2)$. This will remain finite and chain-dependent as the link length tends to zero, so different  perfectly flexible strings will produce different fountains heights. The  rope fountain, with finite bending stiffness rather than links, is also an  open problem.

This paper confirms that when a chain is picked up, part of its momentum comes from the surface it is picked up from. In any industrial or technological setting where a chain is being deployed, accurate predictions about how much force is required will have to include this force. Further work on how to maximize this force may find application. Finally, picking up a chain has traditionally been thought to belong to a wide class of problems in which  half the  work done is dissipated. Charging a capacitor at constant voltage is a better known electrical example. The surprising upwards push during  chain pickup  increases the fraction of energy that is retained  to $1/(2(1-\alpha))$. Perhaps it is worth revisiting other traditional problems in this class to see whether similar effects can be harnessed to reduce energy dissipation.

\end{document}